\documentclass[sigconf]{acmart}

\usepackage{amsmath}

\usepackage{bm}
\usepackage{tabularx}
\usepackage{multirow}
\usepackage{subcaption}
\newcolumntype{L}[1]{>{\raggedright\arraybackslash}p{#1}}
\newcolumntype{C}[1]{>{\centering\arraybackslash}p{#1}}
\newcolumntype{R}[1]{>{\raggedleft\arraybackslash}p{#1}}

\newtheorem{defn}{Definition}

\renewcommand\footnotetextcopyrightpermission[1]{} % added by Yao
\AtBeginDocument{%
  \providecommand\BibTeX{{%
    \normalfont B\kern-0.5em{\scshape i\kern-0.25em b}\kern-0.8em\TeX}}}

\setcopyright{acmcopyright}
\copyrightyear{2018}
\acmYear{2018}
\acmDOI{10.1145/1122445.1122456}

\acmConference[Woodstock '18]{Woodstock '18: ACM Symposium on Neural
  Gaze Detection}{June 03--05, 2018}{Woodstock, NY}
\acmBooktitle{Woodstock '18: ACM Symposium on Neural Gaze Detection,
  June 03--05, 2018, Woodstock, NY}
\acmPrice{15.00}
\acmISBN{978-1-4503-XXXX-X/18/06}

\begin{document}

\pagestyle{plain} 
\title{From Intrinsic to Counterfactual: On the Explainability of Contextualized Recommender Systems}


\author{Yao Zhou}
\authornotemark[1]
\email{yaozhou3@illinois.edu}
\affiliation{%
  \institution{University of Illinois Urbana Champaign}
  \city{Urbana}
  \country{USA}
}

\author{Haonan Wang}
\authornotemark[1]
\email{haonan3@illinois.edu}
\affiliation{%
  \institution{University of Illinois Urbana Champaign}
  \city{Urbana}
  \country{USA}
}

\author{Jingrui He}
\email{jingrui@illinois.edu}
\affiliation{%
  \institution{University of Illinois Urbana Champaign}
  \city{Urbana}
  \country{USA}
}

\author{Haixun Wang}
\email{haixun@gmail.com}
\affiliation{%
  \institution{Instacart}
  \city{San Francisco}
  \country{USA}
}

% \renewcommand{\shortauthors}{Yao and Haonan, et al.}
% \renewcommand{\shortauthors}{Yao and Haonan, et al.}

\begin{abstract}
    With the prevalence of deep learning based embedding approaches, recommender systems have become a proven and indispensable tool in various information filtering applications. However, many of them remain difficult to diagnose what aspects of the deep models' input drive the final ranking decision, thus, they cannot often be understood by human stakeholders. In this paper, we investigate the dilemma between recommendation and explainability, and show that by utilizing the contextual features (e.g., item reviews from users), we can design a series of explainable recommender systems without sacrificing their performance. In particular, we propose three types of explainable recommendation strategies with gradual change of model transparency: whitebox, graybox, and blackbox. Each strategy explains its ranking decisions via different mechanisms: attention weights, adversarial perturbations, and counterfactual perturbations. We apply these explainable models on five real-world data sets under the contextualized setting where users and items have explicit interactions. The empirical results show that our model achieves highly competitive ranking performance, and generates accurate and effective explanations in terms of numerous quantitative metrics and qualitative visualizations. 

%   \he{The abstract is very poorly written.} Recommender systems play an increasingly important role in various information filtering applications, but often in ways that are difficult to understand by human stakeholders, especially with the prevalence of deep learning techniques. By uncovering the user preference using complex modeling, it seems like the diminished explanability is becoming an issue for modern recommender systems. In this paper, we study the dilemma between deep learning based recommendation and explanability, and show that we do not have to sacrifice the performance in order to achieve explanability. In particular, we performe horizontal comparisons among various types of explainable recommendation strategies with gradual change of model transparency (i.e., whitebox, graybox, and blackbox). Specifically, we propose to use three recommendation models to explain their ranking decisions (attention weights, adversarial perturbations, and counterfactual perturbations) under the contextualized setting where users and items have explicit interactions. The empirical experiments show that these three models can achieve highly competitive ranking performance, yet also shed a light on model explanations with promising results w.r.t. user preferences.
\end{abstract}




\maketitle

{
\renewcommand{\thefootnote}{\fnsymbol{footnote}}
\footnotetext[1]{Both authors contributed equally to this work.}
}

\section{Introduction}
Recommendation plays an important role in many information filtering systems, such as e-commerce (Amazon, Walmart), streaming services (YouTube, Netflix), and business review services (Yelp, Trip Advisor), etc. Modern recommender systems (RS) provide personalized recommendation suggestions by learning from the historical user-item interactions, e.g., explicit interactions such as ratings or implicit interactions such as clicks. Recently, with the prevalence of embedding-based approaches, many recommender systems can accurately uncover the preference of users over unseen items. By learning the ranking model merely from the historical interactions data between users and items, a lot of successes have been observed in both industry and academia. Nevertheless, most (if not all) of these models suffer from a lack of model explainability. Over the last few years, the demand for explainability of recommender systems drastically increased because customers are no longer satisfied with only high-quality recommendations, but they also require intuitive explanations. Explaining the relationship, in a human-interpretable way, between the users and its ranking decisions is critical for transparency, effectiveness, and trustworthiness. By providing a personalized recommender algorithm that explains why the recommendation results are presented in such a specific way, could significantly improve user satisfaction.

Without using the explicit user-item interaction features (e.g., review text), the pioneering work of recommender systems that utilized item-based collaborative filtering methods are explainable to some degrees, e.g. ``similar movies other users also watched'' or ``this product is similar to the products you purchased before'', etc. Later on, the content-based recommender system becomes prevalent since it is also human interpretable by modeling users and items with various shared content information, for example, ``genre'', ``actors'', and ``duration'' in the movie recommendation problem. Besides using simple models with good explainability such as regression-based or tree-based models, there exist multiple strategies to build content-based explanation recommender systems. With the success of more business-driven models being applied into productions, such as matrix factorization based models~\cite{PMF_recsys_2008, MF_Netflix_2009}, neural collaborative filtering based models~\cite{NCF_www17, APR_SIGIR_2018}, generative adversarial network based models~\cite{IRGAN_SIGIR_2017, PURE_KDD_2021}, and graph-based models~\cite{lightGCN_2020}, the performance of recommender systems has dramatically improved over the last decade, but the lack of explainability is becoming a more severe concern.

The available solutions to the aforementioned issues mostly rely on incorporating contextualized features into the model learning process. In recent years, multiple efforts have been made in this direction, such as FM~\cite{FM_ICDM_2010}, VBPR~\cite{vbpr_aaai_2016}, DeepFM~\cite{deepfm_ijcai_2017}, etc. The majority of such efforts are devoted to designing the learning mechanism of the high-order feature interactions. Typical contextualized features for users could be their meta-information (e.g., locations, age, gender), and those for items could come from other sources (e.g., images, item titles, item descriptions). However, since their contextualized features are isolated between users and items, these models are more suitable for ``cold start'' recommendations and personalized ranking with rich features. Another line of research, such as EFM~\cite{EFM_2014}, utilized the user-item interaction features (text reviews) and aimed to learn the sentiment latent factors using matrix factorization. However, their learning objective is explicitly modeled for personalized sentiment attention instead of user preference. Therefore, it is difficult to directly apply to personalized ranking problems.

In this work, we propose to achieve the explainability of personalized ranking by building a series of models with various architecture transparency: an intrinsically interpretable model (whitebox model with feature attention); a model with partial transparency (greybox model with adversarial training); and a post-hoc explanations model (blackbox model with counterfactual augmentations). Intuitively, we could easily build a fully transparent model such as logistic regression or decision tree that has good explainability. However, they mostly have limited performance (due to model linearity or inclination to overfitting) under many application scenarios. In this work, we mainly focus on providing insights into the explainability regarding existing neural network based recommendation models and perform a horizontal comparison among all proposed strategies in terms of their characteristics and relationships. 

The main contribution of this paper can be summarized as follows:
\begin{itemize}
    \item We provide a systematic overview of the explainable recommendation models based on the level of model transparency requirement in a descending order: intrinsic attention based recommendation, adversarial explainable recommendation, and counterfactual explainable recommendation.
    \item We compare the advantages and disadvantages of these three types of models in terms of their effectiveness, suitable domains, as well as their explainability via various qualitative visualizations.
    \item We verify our claims and quantitatively evaluate all three models on five publicly accessible data sets in terms of user explanation oriented metrics and user ranking oriented metrics.
\end{itemize}
The rest of the paper is organized as follows. Section 2 is the related work. Section 3 describes the proposed three explainable recommendation models, and Section 4 presents the analysis of these models from various perspectives. The experimental results are demonstrated in Section 5, and we conclude the paper in Section 6.

\section{Related Work}
In this section, we introduce the background knowledge regarding the explainable recommendation and other highly related domains: counterfactual explanation, adversarial ranking, and attention-based recommendation.

\subsection{Explainable Recommendation}
Recommender systems point the online users to a certain list of items with potential interests, thereby increasing the cross-sales and customer engagement. Many efforts have been devoted to this direction with the booming growth of e-commerce and online streaming services. Recent years have seen significant improvement in performance with the emergence of more sophisticated models. However, only showing the recommended ranking list of items can hardly gain the trust and satisfaction of users. The reason is that many modern recommendation models are difficult to support the end-users in the decision-making processes by providing useful and meaningful explanations~\cite{XRS_design_evaluation_2015,XRS_compare_2014,XRS_Survey_Perspective}. Traditional explainable models such as collaborative filtering~\cite{explain_CF_2000,item_CF_2001}, factorization-based model~\cite{EFM_2014,XRS_MTL_2018,XRS_ijcai_2020}, and tree-based models~\cite{TEM_XRS_2018} can leverage the explanations from user/item features~\cite{ERS_content_2019}, opinions~\cite{topic_XRS_2013}, and summarized topics~\cite{PGM_CF_2015}. However, they usually have limited performance and lack the ability to extend to complicated scenarios. Recently, various types of neural network based explainable recommendation models have been proposed. A2CF~\cite{XRS_Personalized_2020} is an attribute-aware model that uses the residual network to model user-item reviews with joint sentiment analysis; SAERS~\cite{XRS_gradcam_2019} leveraged the power of gradient-weighted class activation mapping~\cite{gradCAM} and used the backpropagated ranking loss to generate item image's saliency map as visual explanations; AMCF~\cite{xrs_feature_mapping_2020} maps the uninterpretable features into the interpretable aspect features by minimizing both ranking loss and interpretation loss, etc. Various kinds of explanation models are emerging every year, e.g., set operation based model~\cite{xrs_scrutable_2019}, knowledge graph based explanation model~\cite{KD_XRS_2018}, disentangled embedding explanation~\cite{Disentangle_RS}, and many others.

These existing approaches generate explanations in different ways but lack the systemic analysis regarding explainable recommendation models from multiple perspectives. Based on the classification of explainable models ~\cite{XRS_Survey_Perspective}, they can be either information source dependent (displaying explainable content that is human-readable as the justification of the ranking model's decision) or model architecture-dependent (deliberately designing the machine learning model with the ability to explain). In our work, we have considered both directions and studied three types of models that are coherently connected as well as complying with this classification taxonomy.  

\subsection{Attention Based Recommendation}
Attention models are one type of input processing technique that allows the neural network to focus on certain parts of a complex input by assigning various modular learned weights. In this way, it allows the neural networks to approximate the visual attention mechanism humans use and the output can be calculated as an attention-weighted sum of the input. It has gained a lot of successes in multiple domains especially natural language processing and computer vision. In recent years, various works~\cite{NeuralAttention_WWW_2018, NPA_KDD_2019,NETE_LEI_CIKM_2020,PETER_ACL_2021,XRS-Attetion-aaai-2019} have been proposed to use attention in recommender systems. Intuitively, different users may interact with the same item with attention to different aspects, e.g., words in the text, saliency areas in images, etc. NPA~\cite{NPA_KDD_2019} and NRPA~\cite{NRPA_SIGIR_2019} use two-level attention models of both word-attention and document-attention to personalize news recommendation; NETE~\cite{NETE_LEI_CIKM_2020} learned template-controlled sentences from data to describe word features; VECF~\cite{VECF_SIGIR_2019} utilized both attention-weighted visual features and GRU-based weak supervised learning to increase the model explanation on personalized ranking. In this work, we adopt a similar strategy where the attention weights are learned with ID embeddings and feature embeddings. However, for the sake of explanation, the attentions are applied to features instead of their embeddings.

\subsection{Adversarial Personalized Ranking}
Despite the successes of recommendation models, numerous works have reported that they can be vulnerable when exposed to adversarial perturbations or even random perturbations~\cite{APR_SIGIR_2018,AdvRS_CS_2021}. Thus, two types of adversarial recommendation models have been proposed: adversarial learning~\cite{FGSM_ICLR_2015,PGD_ICLR_2018} based RS and GAN-based RS. Among the first type: APR~\cite{APR_SIGIR_2018} proposed to add adversarial perturbations on user/item embeddings to improve BPR\cite{BPR_rendle_2009} performance; AMR~\cite{AMR_TKDE_2020} leverages the adversarial training by adding adversarial perturbations to target image features. For the second type: IRGAN~\cite{IRGAN_SIGIR_2017}, for the first time, formulated the IR tasks as a minimax game where the generator learns the discrete relevance distribution of entities and the discriminator is the ranking model that decides whether the user-item pair is relevant or not; CFGAN~\cite{CFGAN_CIKM_2018} relaxed the discrete constraints of IRGAN by introducing a generator with the ability to perform continuous embedding space synthesizing; PURE~\cite{PURE_KDD_2021} further enriched the CFGAN by performing a positive-unlabeled sampling strategy and reached state-of-the-art performance. Nevertheless, none of them have the ability to provide explanations for the decisions made by the ranking model. Our work bridges the gap by introducing an explanation to the ranking model via adversarial training. 

\subsection{Counterfactual Explanation}
Counterfactual explanation is a specific class of explanation model that provides the reasoning between decision making and input modifications. In recommendations, counterfactual explanations provide perturbations to user input features and these modifications can help the ranking model's transition to the desirable decision boundary, e.g., flipping the original relevance preference over pair of items. Multiple efforts have been made to categorize and evaluate the counterfactual explanation models~\cite{CF_ML_REVIEW_2020,CF_BB_GDPR_2017} and they have found successful applications in multiple domains, including natural language processing, computer vision~\cite{CF_CV_ICML_2019}, and recommender systems~\cite{CF_RS_cikm_2021,CF_Sequential_sigir_2021,tran2021counterfactual, ghazimatin2020prince}. This is an emerging domain in recent years, and our work extends the existing work with multi-pass counterfactual augmentations.

\section{Proposed Model}

% The explainable recommendation model will not only infer the underlying ranking list of the unseen items for the user, but also output explanations for the ranking decision in a human interpretable way. The pioneer works of recommender systems that utilized item-based collaborative filtering methods are explainable to some degree, e.g. ``similar movies other users also watched'' or ``this product is similar to the products you purchased before'', etc. Later on, the content based recommender system becomes prevalent since it is also human interpretable by modeling users and items with various shared content information, for example, ``genre'', ``actors'', and ``duration'' in the movie recommendation problem. Besides using simple models with good explainability such as regression-based or tree-based model, there exists multiple strategies to build content-based explaination recommender systems. With the success of more business driven models being applied into productions, such as matrix factorization based models, neural collaborative filtering based models, generative adversarial network based models, and graph based models, the performance of recommender systems has dramatically improved over the last decade, but the lack of explainability is becoming a concern to the community especially for the sake of gaining trust and ensuring fairness.\he{Isn't this also part of related work, and NOT the proposed model?} 

\subsection{Problem Definition}
In this paper, we study the contextual recommendation problem where $\mathcal{U}$, $\mathcal{I}$, and $\mathcal{F}$ are the set of users, items, and features, respectively. Given a user $u \in \mathcal{U}$, a list of relevant items and their corresponding interactions are recorded. For each user-item pair $(u, i)$, there exists a group of raw features that can be used as the source of explanation to describe the interaction between the user $u$ and the item $i$. A good source of features is the review text from the user for the specific items. Compared with many existing works that only utilized the isolated features from either the user (e.g., biographical information) or the item (e.g., item title description or item image), the text review features have the explicit interactions between user and item, thus, it is more informative in terms of building an explainable ranking model. We further assume $\Omega$ to be the index set of observed user-item-feature interactions $(u, i, \bm{f}_{u,i}) \in \Omega$ where we denote $\bm{f}_{u,i}$ as the observed review features within one specific application domain (e.g., electronics, fashion). Here, the corresponding user id, item id, as well as their review features should satisfy $u \in \mathcal{U}, i \in \mathcal{I}$ and $\bm{f}_{u,i} \subset \mathcal{F}$.

Then, our recommendation problem is formulated as follows: 
\begin{defn}
[\textbf{Contextualized Explainable Recommendation}] \leavevmode \\
\textbf{Given:} A set of users $\mathcal{U}=\{u_1, u_2, ..., u_M\}$, a set of items $\mathcal{I}=\{i_1, i_2, ..., i_N\}$, a set of raw features $\mathcal{F}=\{w_1, w_2, ..., w_P\}$, the observed user-item-feature interaction sets $\Omega$. \\
\textbf{Output:} The estimated interaction scores of the unobserved items for each user $u$ in $\mathcal{U}$ as well as their corresponding feature explanations.
\end{defn}

It should be noted that under this problems setting, the features being utilized for users are not isolated from these ones from items, this is different from the traditional contextualized recommender systems~\cite{deepfm_ijcai_2017,vbpr_aaai_2016,FM_ICDM_2010}. Similar to~\cite{EFM_2014, CF_RS_cikm_2021}, we decompose the review features $\bm{f}_{u,i} \in \mathbb{R}^{p}$ (where $p < P$) into user feature vector $\bm{f}_u \in \mathbb{R}^{p}$ and item feature vector $\bm{f}_i \in \mathbb{R}^{p}$ w.r.t. the $k$-th review feature $w_k$ as shown in Equation~(\ref{eq_feature_preprocess}). Intuitively, the user tends to comment on the features they care more about, thus, the user feature is merely a frequency-based vector. However, for items, their qualities can be accurately reflected in the review sentiments from multiple users. Therefore, we formulate the item features by considering both review feature frequency and average sentiment.
\begin{equation}
    \begin{split}
        \bm{f}_u^k &= 1 + (T-1) \Bigg( \frac{2}{1+\mathrm{exp}\left(-\bm{f}_{u,i}^k\right)} - 1 \Bigg)\\
        \bm{f}_i^k &= 1 + \frac{T-1}{ 1 + \mathrm{exp}\left(- \bm{f}_{u,i}^k \cdot s_i^k \right) }
    \end{split}
    \label{eq_feature_preprocess}
\end{equation}
where $T$ is the maximum review scores a user can give to an item, $\bm{f}_{u,i}^k$ is the frequency that user $u$ mentioned review feature $k$ toward item $i$, and $s_i^k$ is the average sentiment on review feature $k$ toward item $i$. After transformation, all entries of these feature vectors are mapped into the range of $[1, T]$ and the the indexed set becomes $\Omega = \{ (u, i, \bm{f}_u, \bm{f}_i) \}$.

\begin{figure}[!t]
\centering
\includegraphics[width=8.0cm]{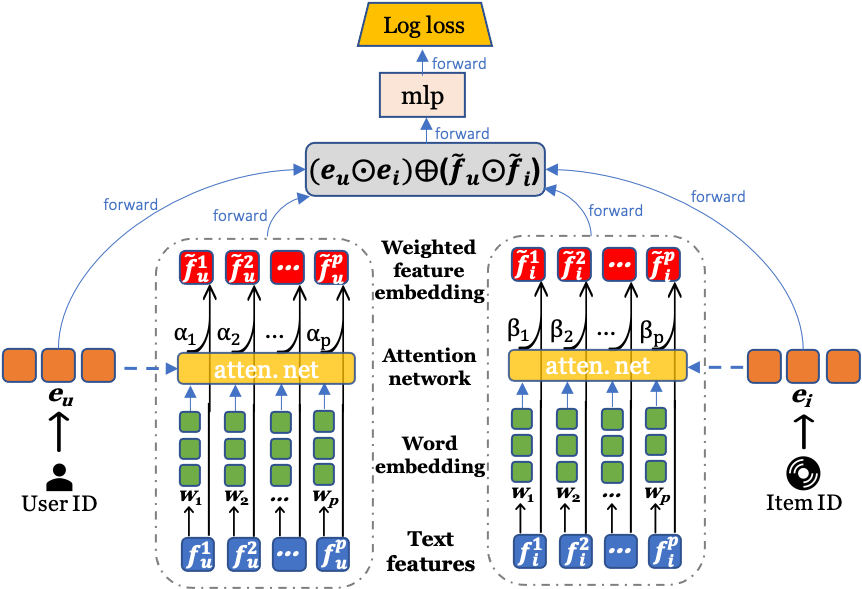}
\vspace{-2mm}
\caption{Attention based model: NAR}
\label{attention-model-NAR}
\vspace{-2mm}
\end{figure}

\subsection{Intrinsic Explanation}
In this subsection, we introduce the neural attention recommendation (NAR) in detail. NAR is a white box recommendation model with intrinsically explainable designs and it has two major components: a user (item) network to learn user (item) representations, and a final prediction network to predict the relevance scores. The overview of the NRA model is shown in Figure~\ref{attention-model-NAR}. 

\subsubsection{User (item) embedding network} For the embedding network, it maps each user $u$ and item $i$ into lower-dimensional representations $\bm{e}_u \in \mathbb{R}^{m}$ and $\bm{e}_i \in \mathbb{R}^{m}$ via an embedding layer based on their IDs. Under the contextualized setting, we also have the decomposed feature vectors $\bm{f}_{u}$ and $\bm{f}_{i}$ available for each user $u$ and each item $i$. Given one feature vector $\bm{f}_{u}$ or $\bm{f}_{i}$, we embed each word $w_k$ within it to a lower dimensional representation $\bm{w}_k \in \mathbb{R}^{n}$ (denoted in bold) via word embeddings.
%\he{The notation here is very confusing. Also, you mentioned in the last section that `we decompose the review features $\bm{f}_{u,i}$ into user feature vector $\bm{f}_u$ and item feature vector $\bm{f}_i$ w.r.t. the $k$-th review feature $w_k$'.}

Within a specific application domain, each user will have attentions on a few features. Based on this intuition, we highlight the crucial features by designing the attention network as:
\begin{equation}
    \alpha_k = \frac{ \mathrm{exp}(\bm{e}_{u}M_a \bm{w}_k) }{ \sum_{k=1}^{p} \mathrm{exp}(\bm{e}_{u}M_a \bm{w}_k)}    
\end{equation}
where $M_a \in \mathbb{R}^{m \times n}$ is the attention mapping matrix between user embedding and word embedding. Thus, the attention weights of each user on $k$-th word will be $\alpha_k \in (0,1)$. Each item can also obtain its attention weights $\beta_k \in (0,1)$ in a similar manner with a different attention mapping matrix $M_b \in \mathbb{R}^{m \times n}$.  

\subsubsection{Prediction network}
To incorporate the review features into the prediction, one naive solution is to let the user and item share the same feature vector $\bm{f}_{u,i}$. In our design, different mapping functions, i.e., Equation~(\ref{eq_feature_preprocess}), are applied to transform $\bm{f}_{u,i}$ to the user as $\bm{f}_u$ and to the item as $\bm{f}_i$, respectively. After the embedding network, we obtain the attention aggregated representations of users and items as follows. 
\begin{equation}
    \Tilde{\bm{f}}_u = \sum_{k=1}^{p} \alpha_k \bm{w}_k \bm{f}_u^k \quad \mathrm{and} \quad \Tilde{\bm{f}}_i = \sum_{k=1}^{p} \beta_k \bm{w}_k \bm{f}_i^k
\end{equation}
In the end, we concatenate the user and item embeddings and attention aggregated representations. Then, the final relevance prediction between the user and item can be retrieved by passing the concatenated vector representation into a multi-layer perceptron (MLP) layer\footnote{We could also use factorization machine to learn the high-order interactions, but without loss of generality, we adopt MLP as the final layer for all baselines to assure an easy and fair comparison between various models.}
\begin{equation}
    \mathrm{R}_{u,i}^{\mathrm{NAR}} = {\tt MLP} \Big( (\bm{e}_u \odot \bm{e}_i) \oplus (\Tilde{\bm{f}}_u \odot \Tilde{\bm{f}}_i) \Big)
\end{equation}
where $\oplus$ is the concatenation of two vectors and $\odot$ is the element-wise product of two vectors.

\begin{figure}[!t]
\centering
\includegraphics[width=8.0cm]{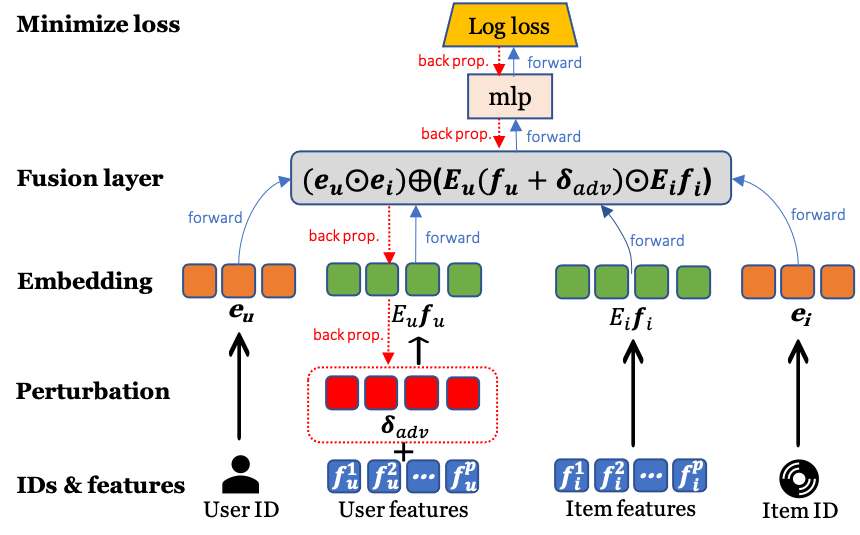}
\vspace{-2mm}
\caption{Adversarial based model: CAR}
\label{adversarial-model-CAR}
\vspace{-2mm}
\end{figure}

\subsection{Adversarial Explanation}
In this section, we introduce the contextualized adversarial ranking model (CAR) in detail. CAR is a gray box recommendation model with a partial transparency requirement on the feature gradients and it allows the ranking model to be retrained with adversarial augmentations. Specifically, CAR has the same user (item) embedding network as NAR. In addition, CAR also has two distinct components: an adversarial augmentation network, and the final prediction network. The overview of the NRA model is shown in Figure~\ref{adversarial-model-CAR}. 

% \subsubsection{User (item) embedding network}
% Similar to NAR, as the first step, CAR also maps each user $u$ and item $i$ into lower-dimensional representations $\bm{e}_u$ and $\bm{e}_i$ based on their IDs. Meanwhile, the feature input $\bm{f}_{u,i}$ is also mapped into user feature vector and item feature vector respectively w.r.t. $k$-th review feature $w_k$:
% \begin{equation}
%     \begin{split}
%         \bm{f}_u^k &= 1 + (T-1) \Bigg( \frac{2}{1+\mathrm{exp}\left(-\bm{f}_{u,i}^k\right)} - 1 \Bigg)\\
%         \bm{f}_i^k &= 1 + \frac{T-1}{ 1 + \mathrm{exp}\left(- \bm{f}_{u,i}^k \cdot s_i^k \right) }
%     \end{split}
%     \label{eq_feature_preprocess}
% \end{equation}
% where $T$ is the maximum review scores an user can give to an item, $\bm{f}_{u,i}^k$ is the frequency that user $u$ mentioned review feature $k$ toward item $i$, and $s_i^k$ is the average sentiment on review feature $k$ toward item $i$.

\subsubsection{Prediction network}
To incorporate the feature signal into the ranking prediction, a common practice is first to convert user features and item features to latent embeddings, then, infuse them with the user (item) ID embeddings as a single latent vector representation. 
\begin{equation}
    \mathrm{R}_{u,i}^{\mathrm{CAR}} = {\tt MLP} \Big( (\bm{e}_u \odot \bm{e}_i) \oplus (E_u\bm{f}_u \odot E_i\bm{f}_i) \Big)
\end{equation}
where $E_u$ and $E_i$ are the mapping matrices that convert the normalized user feature vector and item feature vector into the same latent space.

\subsubsection{Adversarial augmentation network}
Considering that we aim to use adversarial perturbation for generating the explainable perturbations, the first step is to learn the base recommendation model\footnote{For a fair comparison, all baselines are modified to use the pairwise loss on top of the neural Bayesian personalized ranking (BPR) architecture and all optimization hyper-parameters are set to be the same, e.g., learning rate, embedding size.}. Then, we generate the adversarial perturbations that aim to maximize the loss of our model. Assuming that the BPR loss is $L_{\mathrm{BPR}}=-\sum_{(u,i,j)} \mathrm{log}[\sigma(R_{u,i}^{CAR}-R_{u,j}^{CAR})]$, then, the corresponding perturbations could be generated using the fast gradient sign method (FGSM~\cite{FGSM_ICLR_2015}):
\begin{equation}
    \bm{\delta}_{adv} = \epsilon \frac{\Delta}{||\Delta||_{2}} \quad \mathrm{where} \quad  \Delta = \frac{ \partial L_{\mathrm{BPR}} }{ \partial {\bm{f}_u} }
\end{equation}
where $\Delta$ is the perturbation estimated by calculating the gradient of the BPR loss w.r.t. the user feature vectors. This assumes that our loss is the first-order derivative around $\bm{f}_u$. Meanwhile, the adversarial perturbation is under the max-norm constraint. In other words, after the L2 normalization, each dimension of $\bm{\delta}$ should be constrained under the magnitude of $\epsilon$ in order to get a scale-controllable minimum perturbation. 

Next, we augment the original training set with adversarial perturbations on top of the user features. Notice that adversarial augmentation with retraining could in general make the ranking model less sensitive to the adversarial perturbations, hence increasing the model robustness. In our contextualized setting, we could augment the data set with item features or even user (item) id embedding as well. However, as the target is to provide explanations for personalized rankings, we only need to perturb the user feature vector to serve this purpose. Eventually, our adversarial augmented prediction network will be trained in the following manner:
\begin{equation}
    \mathrm{R}_{u,i}^{\mathrm{CAR_{adv}}} = {\tt MLP} \Big( (\bm{e}_u \odot \bm{e}_i) \oplus \Big( E_u (\bm{f}_u + \bm{\delta}_{adv}) \odot E_i\bm{f}_i \Big) \Big)
\end{equation}

From another perspective, we can think of the loss of the adversarial augmentations as a regularization term to control the trade-off between the ranking model that optimizes the pairwise loss and the attacking model that finds the most effective perturbations against the current ranking model. This trade-off can be balanced by introducing a hyper-parameter $\lambda$:
\begin{equation}
    \min_{\theta} L_{BPR}(\Omega|\theta) + \lambda L_{BPR}(\Omega_{adv}|\theta)   
\end{equation}
where $\theta$ is the ranking model's parameter and $\Omega_{adv}$ is the user-item-feature set where the user feature vector has its adversarial perturbation $\delta_{adv}$ being added on user features.

\begin{figure}[!t]
\centering
\includegraphics[width=8.0cm]{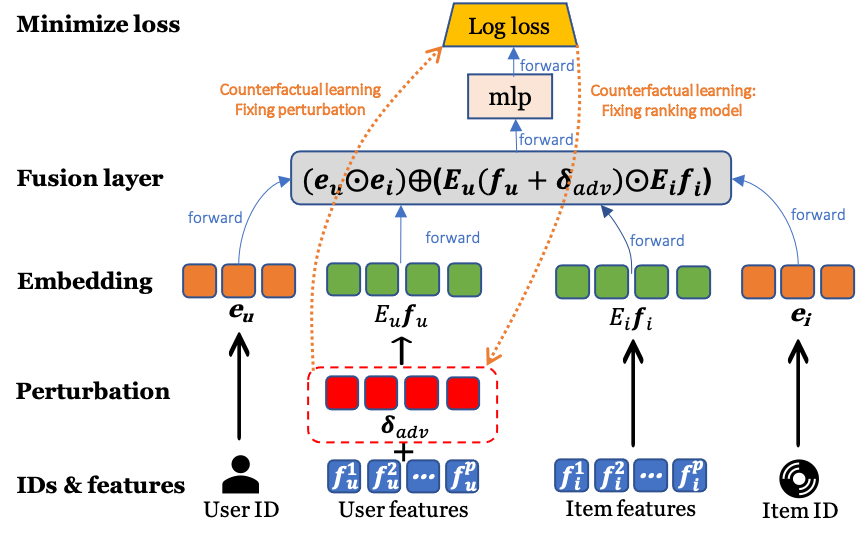}
\vspace{-2mm}
\caption{Counterfactual model: CNR}
\label{adversarial-model-CNR}
\vspace{-2mm}
\end{figure}

\subsection{Counterfactual Explanation}
In this section, we introduce the counterfactual neural recommendation model (CNR) in detail. CNR is a blackbox recommendation model with model-agnostic designs to generate explanations for rankings. Similar to the CAR model, CNR has exactly the same user (item) embedding network and final prediction network. Nevertheless, its augmentation network is designed by following the counterfactual thinking~\cite{CF_RS_cikm_2021,CF_XRS_A_Review}: {\tt ``User X purchased item A over item B because feature variables have values} ($f_1, f_2, ...$) {\tt associated with X. If user X has a few feature variables} ($f_1', f_2', ...$) {\tt changed and all other variables remained the same, user X would purchase item B over item A instead''}.

One should note that many such explanations exist, our objective is to find a counterfactual perturbation that alters features as little as possible but returns the closest world where the ranking decision has been flipped. Based on the above principle, we can find the counterfactual perturbation $\bm{\delta}_{cf}$ and an optimum solution $\theta$ for the ranking model by solving the following minimax problem:
\begin{equation}
    (\bm{\delta}_{cf}, \theta) = \mathrm{arg} \min_{\delta_{cf}} \max_{\theta} \;  L_{BPR}(\Omega_{cf}|\theta) + \xi \, \mathrm{dist}(\bm{f}_u, \bm{f}_u + \bm{\delta}_{cf})
    \label{prob:counterfactual}
\end{equation}
where $\bm{\delta}_{cf}$ is the counterfactual perturbation of user features by fixing the model parameter $\theta$ and minimizing this objective. $d(\cdot, \cdot)$ is the distance function that measures the difference between original user features and perturbed user features. Meanwhile, the ranking loss should be maximized since the ranking decision is designed to be flipped after the feature perturbations are applied. In practice, we should initialize $\theta$ by training the ranking model without any perturbations, and then we can iteratively update $\bm{\delta}_{cf}$ and $\theta$ until a perturbation has been found that is sufficiently small to flip the ranking predictions. For example, under the pairwise learning setting, $R_{u,i} > R_{u,j}$ will become $R_{u,i} < R_{u,j}$ if the counterfactual perturbation is applied on user features $\bm{f}_u + \bm{\delta}_{cf}$.

The distance term $\mathrm{dist}(\cdot, \cdot)$ is critical in counterfactual recommendations, especially when we aim to generate human interpretable explanations. Namely, only a small number of features should be changed and the rest remain untouched. In this way, these counterfactual perturbations are easier for humans to interpret. Thus, various distance functions can be defined to support our goal. One straightforward definition is L2 distance $\mathrm{dist} = ||\bm{\delta}_{cf}||_2^2$, which can be used to guarantee the user features are minimally perturbed. For the sake of introducing sparsity to $\bm{\delta}_{cf}$ so that humans can understand such explanation, we also introduce the L1 norm $\mathrm{dist} = ||\bm{\delta}_{cf}||_2^2 + ||\bm{\delta}_{cf}||_1$. This is also called elastic net regularization, which combines ridge and lasso regularization. 

% Another way to introducing sparsity is applying mask to perturbation where we constrain the perturbations can only happened on a few pre-selected index position $dist = ||\bm{\delta}_{cf} \odot \bm{\pi}||_2$ where $\bm{\pi}$ is the corresponding mask vector. This is an appealing property because we can incorporate the prior information of the mask vector $\bm{\pi}$. For example, we could initialize $\bm{\pi}$ as the explanation vectors from the intrinsic model, adversarial model, or any prior indicators.

\begin{table*}[!t]
\centering
% \small
\setlength\tabcolsep{3pt}
\begin{tabular}{l|c|c|c|c|c}
\toprule
& \textbf{Model Transparency} & \textbf{Data Augmentation} & \textbf{Constraint} & \textbf{Learning Objective} &\textbf{Extensibility} \\\midrule
\textbf{Intrinsic Model}      & Fully          & False          & ---     & Minimize log loss & Low               \\ \midrule
\textbf{Adversarial Model}    & Partial        & True           & Max-norm bounded       & Maximize loss gradient   & High              \\ \midrule
\textbf{Counterfactual Model} & Model agnostic & True           & P-norm regularized      & Flip ranking decision   & High           \\ 
\bottomrule
\end{tabular}
\caption{Comparison of three types of explanation models}
\label{XRS_comparions_table}
\vspace{-6mm}
\end{table*}

\section{Model Analyses}
We compare all three types of explanation models in various aspects: model transparency, data manipulation, explanation constraints, learning objectives, and extensibility. Specially they all share certain commonalities but have a significant difference in terms of their explanation generating mechanisms. The high-level summary is presented in Table~\ref{XRS_comparions_table}.

\subsubsection*{Model transparency}
From the intrinsic model to the adversarial model and then the counterfactual model, the requirements for model transparency gradually decrease. The intrinsic model (NAR) usually requires full transparency of the ranking model architecture and a well-designed attention mechanism with extra attention weights on user features. The adversarial model (CAR) only requires the gradient w.r.t. the features that need to be explained, and it is more like a partially transparent model. The counterfactual model (CNR) bypasses the substantial challenges of exploring the internal logic of the complex ranking systems. Therefore, counterfactual models can explain the rationale of the ranking decision-making process to a certain extent without opening the ``black box'', and it is a model agnostic explanation approach.

\subsubsection*{Learning objectives}
The learning objectives of all three types of models will be the same as any popular ranking models where we usually minimize the log loss of the observed user-item interactions regardless of whether it is under the pointwise or the pairwise setting. The major difference is reflected in their explanation components: for the intrinsic model, the explanation is instantiated using attention weights, thus, no modification is needed in the objective; for the adversarial model, the perturbation is designed to be the gradient w.r.t. the user features that can maximize the log loss; for the counterfactual model, the perturbation is learned to serve the purpose of flipping the pre-trained ranking model's decision, do it does not necessarily need to be the largest gradient w.r.t. the log loss. 

\subsubsection*{Explanation constraints}
The intrinsic model has no constraints on the learning process. The adversarial model's perturbation is max-norm bounded by the magnitude of $\epsilon$ on the loss's gradient. The counterfactual model's perturbation is usually L2-norm regularized (and L1-norm regularized sometimes for the sake of explanation sparsity) but not strictly bounded. 

\subsubsection*{Data manipulations}
The intrinsic model does not require any data augmentations since it is inherently designed to be explainable by attention weights. Both the adversarial model and counterfactual model require data augmentation to increase the model expressiveness and alleviate the data sparsity issue by adding more data points near the ranking decision boundary in latent feature space.

\subsubsection*{Extensibility}
The adversarial model and counterfactual model require very little or no effort to modify the existing model architectures. Therefore, they can be easily extended from any contextualized recommender systems. On the other hand, the intrinsic model requires a dedicated design for the attention mechanism.

\section{Experimental Results}
In this section, we aim to answer the following questions: \\
\noindent \textbf{RQ1:} Do the learned models give competitive and human interpretable results? \\
\noindent \textbf{RQ2:} Are the explainable mechanisms helpful with the ranking performance? \\
\noindent \textbf{RQ3:} How do we evaluate the explanation results both quantitatively and qualitatively?

\begin{table}[h!]
\setlength\tabcolsep{1pt} % default value: 6pt
\centering
\scalebox{0.95}{
\begin{tabular}{l|c|c|c|c|c}
\toprule
\textbf{Dataset} & \multicolumn{1}{l|}{\textbf{\# Users}} & \multicolumn{1}{l|}{\textbf{\# Items}} & \multicolumn{1}{l|}{\textbf{\# Interactions}} & \multicolumn{1}{l|}{\textbf{Density}} & \multicolumn{1}{l}{\textbf{\# Features}}\\ \midrule
\textbf{Instrument}     & 1,276       & 843       &3,581        & 0.33\% & 325  \\
\textbf{Video}   & 4,333       & 1,486        & 11,759          & 0.18\%  & 350     \\ 
\textbf{Music}     & 5,339       & 3,538       & 40,315       &  0.21\% &  1,366    \\ 
\textbf{Beauty}     & 21,472       & 11,897       & 105,659       & 0.04\% & 1,985    \\ 
\textbf{Clothing}       &  37,703         &  22,647       & 142,553          & 0.02\% & 1,462     \\ \bottomrule
\end{tabular}
}
\caption{Statistics of the data sets}
\vspace{-8mm}
\label{dataset}
\end{table}

\subsection{Experiment Settings}
\subsubsection{Data sets}
We conduct the experiments on five publicly accessible data sets\footnote{ \url{http://jmcauley.ucsd.edu/data/amazon/index_2014.html}} with various sizes: Instrument, Video, Music, Beauty, and Clothing. For the preprocessing steps, we keep the user-item interaction pairs that have text reviews. To perform the negative sampling, we treat the 4-star and 5-star reviews as the positive feedback and the rest are unlabeled feedback~\cite{IRGAN_SIGIR_2017,PURE_KDD_2021}. Following this pro-processing protocol, all user and item interactions are eventually stored into the interaction matrix with entries of values 0 and 1. The details of all data set are summarized in Table~\ref{dataset}. We utilize the Sentire\footnote{\url{https://github.com/lileipisces/Sentires-Guide}} package to further process each data set in order to get the word and sentiment pairs $\{(w_1, s_1), ..., (w_p, s_p)\}$ for each user from the review text.

\subsubsection{Evaluation protocols}
For the purpose of evaluation, we have adopted the $4:1$ train and test random split on each user when the number of positive examples is larger than 5. Otherwise, there will be no positive examples in testing. Similar to the existing work, we also perform the sampled evaluation~\cite{KDD20_SAMPLED_evaluation_RS,NCF_www17} to speed up the computation where each positive test item will be randomly assigned with a candidate item pool of size 100. The final performance evaluation uses both classification and ranking-based metrics, i.e., Precision, Recall, F1, Hit rate, NDCG, and MRR. Specifically, MRR has the cutting threshold of 1 (i.e., MRR@1) and all the rest of the evaluation metrics have the cutting threshold of 10 (e.g., NDCG@10).  

It is also very important that we can quantitatively evaluate the explanations generated by our models. Therefore, besides the model-based performance evaluation, we also perform the user-based evaluation. Intuitively, we aim to evaluate how good our model's explanation is when matched with users' sentiment in terms of the features. Let us assume that user $u$ has $k$ ground truth features $\bm{g}_{u} = \{ g_{u,1}, g_{u,2}, ..., g_{u,k}\}$. Here, a feature will be added to $\bm{g}_{u}$ only if user $u$ has already given sentiments (positive and negative) on this feature regarding all items. Meanwhile, for our ranking model, we have generated an explanation vector $\bm{\phi}_{u} = \{ \delta_{u,1}, \delta_{u,2}, ... \delta_{u,P} \}$, i.e., attention weight vector in NAR or perturbation vector in CAR and CNR. We can sort this explanation vector $\bm{\phi}_{u}$ in descending order and compare with users' sentiment vector $\bm{g}_{u}$ in terms of precision and recall at different cutting thresholds $k$.
\begin{equation}
    \mathrm{Precision}@k = \frac{ |\bm{\phi}_{u,:k} \cap \bm{g}_{u}|}{ |\bm{\phi}_{u,:k}| } \quad \mathrm{and} \quad \mathrm{Recall}@k = \frac{ |\bm{\phi}_{u,:k} \cap \bm{g}_{u}|}{ |\bm{g}_{u}| }
\end{equation}
Furthermore, F1@k is computed on top of precision and recall and NDCG@k will be straight forward to compute as well.

\begin{table*}[!t]
\setlength\tabcolsep{6pt} % default value: 6pt
\centering
\begin{tabular}{l|l|l|l|l|l|l|l|l|l|l}
\toprule
\multirow{2}{*}{} & \multicolumn{2}{c|}{\textbf{Instrument}} & \multicolumn{2}{c|}{\textbf{Video}} & \multicolumn{2}{c|}{\textbf{Music}} & \multicolumn{2}{c|}{\textbf{Beauty}} & \multicolumn{2}{c}{\textbf{Clothing}} \\ \cmidrule{2-11} 
& \textit{F1}            & \textit{NDCG}           & \textit{F1}            & \textit{NDCG}       & \textit{F1}            & \textit{NDCG}   & \textit{F1}            & \textit{NDCG}        & \textit{F1}            & \textit{NDCG}         \\ \midrule
\textbf{NAR} & 0.2028 & 0.3124 & 0.2106 & 0.2991 & 0.3145 & 0.5585  & 0.2286 & 0.3416 & 0.2365  & 0.3302 \\ 
\textbf{CAR} & 0.0160 & 0.0252  & 0.0129 & 0.0220  & 0.0147 & 0.0354  & 0.0163 & 0.0257 & 0.0150 & 0.0200 \\ 
\textbf{CNR} & 0.1858  & 0.2580  & 0.2204 & 0.3417   & 0.3437  & 0.5821 & 0.2279 & 0.3336  & 0.1889 & 0.2535  \\ \bottomrule
\end{tabular}
\caption{User oriented explanation evaluation using classification and ranking metrics}
\label{results:visual-evaluation}
\vspace{-5mm}
\end{table*}

\subsubsection{Baselines}
We have considered various state-of-the-art baselines for the model performance evaluations:
\begin{itemize}
    \item \textbf{Neural collaborative filtering (NCF)}~\cite{NCF_www17}: The most popular two tower based neural network recommendations. 
    \item \textbf{Visual Bayesian personalized ranking (VBPR)}~\cite{vbpr_aaai_2016}: A modified version of the original VBPR by using the neural network based embeddings along with the contextual features. 
    \item \textbf{Counterfactual explainable recommendation (CER)}~\cite{CF_RS_cikm_2021}: The latest counterfactual explainable recommendation model. 
    \item \textbf{Neural attention recommendation (NAR)}: Our proposed attention based intrinsic recommendation model.
    \item \textbf{Contextualized adversarial recommendation (CAR)}: Our proposed personalized ranking model with adversarial explanation. 
    \item \textbf{Counterfactual neural recommendation (CNR)}: Our proposed iterative counterfactual explainable recommendation model.
\end{itemize}

\subsubsection{Reproducible setting}
To guarantee a fair comparison, we fix both ID embedding and feature embedding to be 350 for all models on all data sets. Meanwhile, all models' feature input regarding users and items will be kept the same across different models. We optimize ball baselines using Adam with a fixed learning rate of 0.001 and the number of training epochs is set to 50. For CNR, the number of the outer loop iteration for the minimax problem~(\ref{prob:counterfactual}) is set to 20. For the regularization parameters $\lambda = 1$ for CAR and $\xi = 0.001$ for CNR. The source code (pre-processing and modeling) will be released upon acceptance.

\begin{table}[!h]
\setlength\tabcolsep{1pt} % default value: 6pt
\centering
\scalebox{0.85}{
\begin{tabular}{c|C{13mm}|C{13mm}|C{13mm}|C{13mm}|C{13mm}|C{13mm}}
\toprule
\textbf{Instrument}   & \textbf{Precision} & \textbf{Recall} & \textbf{F1} & \textbf{Hit Rate} & \textbf{NDCG} & \textbf{MRR} \\ \midrule
\textbf{NCF}            & 0.0316 & 0.3047 & 0.0570 & 0.3113 & 0.1773 & 0.1405 \\
\textbf{VBPR}            & 0.0418 & 0.4006 & 0.0753 & 0.4088 & 0.2344 & 0.1871 \\
\textbf{CER}           & 0.0419 & 0.3984 & 0.0753 & 0.4088 & 0.2342 & 0.1881 \\ \midrule
\textbf{NAR}            & 0.0284 & 0.2724 & 0.0513 & 0.2799 & 0.1545 & 0.1206  \\ 
\textbf{CAR}            & \textbf{0.0457} & \textbf{0.4355} & \textbf{0.0822} & \textbf{0.4455} & \textbf{0.2538} & \textbf{0.2025} \\
\textbf{CNR}            & 0.0428 & 0.4077 & 0.0769 & 0.4172 & 0.2406 & 0.1922 \\\bottomrule
\end{tabular}
}
\caption{Model evaluation results of Instrument data set}
\label{results:Musical-Instruments}
\vspace{-5mm}
\end{table}

\begin{table}[!h]
\setlength\tabcolsep{1pt} % default value: 6pt
\centering
\scalebox{0.88}{
\begin{tabular}{C{13mm}|C{13mm}|C{13mm}|C{13mm}|C{13mm}|C{13mm}|C{13mm}}
\toprule
\textbf{Video}   & \textbf{Precision} & \textbf{Recall} & \textbf{F1} & \textbf{Hit Rate} & \textbf{NDCG} & \textbf{MRR} \\ \midrule
\textbf{NCF}            &  0.0535 & 0.5106 & 0.0962 & 0.5222 & 0.2986 & 0.2356 \\
\textbf{VBPR}            & 0.0595 & 0.5638 & 0.1068 & 0.5773 & 0.3362 & 0.2668    \\
\textbf{CER}           & 0.0576 & 0.5482 & 0.1035 & 0.5613 & 0.3226 & 0.2561 \\ \midrule
\textbf{NAR}            & 0.0511 & 0.4893 & 0.0920 & 0.5005 & 0.2875 & 0.2278  \\ 
\textbf{CAR}            & \textbf{0.0596} & \textbf{0.5662} & \textbf{0.1071} & \textbf{0.5800} & \textbf{0.3342} & \textbf{0.2701} \\
\textbf{CNR}            & 0.0579 & 0.5519 & 0.1041 & 0.5666 & 0.3275 & 0.2626  \\\bottomrule
\end{tabular}
}
\caption{Model evaluation results of Video data set}
\label{results:Instant-Video}
\vspace{-5mm}
\end{table}

\begin{table}[!h]
\setlength\tabcolsep{1pt} % default value: 6pt
\centering
\scalebox{0.88}{
\begin{tabular}{C{13mm}|C{13mm}|C{13mm}|C{13mm}|C{13mm}|C{13mm}|C{13mm}}
\toprule
\textbf{Music}   & \textbf{Precision} & \textbf{Recall} & \textbf{F1} & \textbf{Hit Rate} & \textbf{NDCG} & \textbf{MRR} \\ \midrule
\textbf{NCF}            & 0.0628 & 0.3857 & 0.0983 & 0.4753 & 0.2314 & 0.2104 \\
\textbf{VBPR}             & 0.1037 & 0.6300 & 0.1622 & 0.7001 & 0.4520 & 0.4288 \\
\textbf{CER}          & 0.0941 & 0.5634 & 0.1463 & 0.6416 & 0.3843 & 0.3609  \\ \midrule
\textbf{NAR}            & 0.1006 & 0.6106 & 0.1573 & 0.6841 & 0.3989 & 0.3614  \\ 
\textbf{CAR}           & \textbf{0.1081} & \textbf{0.6567} & \textbf{0.1693} & \textbf{0.7244} & \textbf{0.4796} & \textbf{0.4567} \\
\textbf{CNR}    & 0.0974 & 0.5821 & 0.1515 & 0.6584 & 0.4176 & 0.4021       \\\bottomrule
\end{tabular}
}
\caption{Model evaluation results of Music data set}
\label{results:Music}
\vspace{-5mm}
\end{table}

\begin{table}[!h]
\setlength\tabcolsep{1pt} % default value: 6pt
\centering
\scalebox{0.88}{
\begin{tabular}{C{13mm}|C{13mm}|C{13mm}|C{13mm}|C{13mm}|C{13mm}|C{13mm}}
\toprule
\textbf{Beauty}   & \textbf{Precision} & \textbf{Recall} & \textbf{F1} & \textbf{Hit Rate} & \textbf{NDCG} & \textbf{MRR} \\ \midrule
\textbf{NCF}            & 0.0534 & 0.3737 & 0.0885 & 0.4112 & 0.2214 & 0.1881 \\
\textbf{VBPR}            & 0.0775 & 0.5758 & 0.1301 & 0.6074 & 0.3778 & 0.3293 \\
\textbf{CER}         & 0.0665 & 0.4794 & 0.1107 & 0.5110 & 0.3159 & 0.2778   \\ \midrule
\textbf{NAR}            & 0.0617 & 0.4358 & 0.1020 & 0.4652 & 0.2820 & 0.2465  \\ 
\textbf{CAR}            & \textbf{0.0778} & \textbf{0.5791} & \textbf{0.1306} & \textbf{0.6094} & \textbf{0.3776} & \textbf{0.3283} \\
\textbf{CNR}     & 0.0734 & 0.5324 & 0.1221 & 0.5642 & 0.3547 & 0.3121       \\\bottomrule
\end{tabular}
}
\caption{Model evaluation results of Beauty data set}
\label{results:Beauty}
\vspace{-5mm}
\end{table}

\begin{table}[!h]
\setlength\tabcolsep{1pt} % default value: 6pt
\centering
\scalebox{0.88}{
\begin{tabular}{C{13mm}|C{13mm}|C{13mm}|C{13mm}|C{13mm}|C{13mm}|C{13mm}}
\toprule
\textbf{Clothing}   & \textbf{Precision} & \textbf{Recall} & \textbf{F1} & \textbf{Hit Rate} & \textbf{NDCG} & \textbf{MRR} \\ \midrule
\textbf{NCF}            & 0.0364 & 0.3305 & 0.0650 & 0.3504 & 0.1928 & 0.1569        \\
\textbf{VBPR}           & 0.0564 & 0.5252 & 0.1012 & 0.5440 & 0.3112 & 0.2524      \\
\textbf{CER}           & 0.0479  & 0.4409  & 0.0856  & 0.4598  & 0.2665  & 0.2192       \\ \midrule
\textbf{NAR}            & 0.0403 & 0.3707 & 0.0719 & 0.3888 & 0.2146 & 0.1732      \\ 
\textbf{CAR}            & \textbf{0.0578} & \textbf{0.5373} & \textbf{0.1035} & \textbf{0.5562} & \textbf{0.3210} & \textbf{0.2613} \\
\textbf{CNR}       & 0.0513 & 0.4715 & 0.0917 & 0.4920 & 0.2879 & 0.2387            \\\bottomrule
\end{tabular}
}
\caption{Model evaluation results of Clothing data set}
\label{results:Clothin-Shoes-Jewelry}
\vspace{-5mm}
\end{table}

\subsection{Performance Evaluation}
For the recommendation evaluation, the comparison results are summarized in Tables~\ref{results:Musical-Instruments}-\ref{results:Clothin-Shoes-Jewelry}. We observe that regardless of the data set size or the application domain, the adversarial perturbation and augmentation based method CAR always outperforms all other baselines. Meanwhile, we also observe that the counterfactual-based models, such as CER and CNR, also have competitive results. Specifically, CNR consistently beats CER in every aspect. This is because CNR is the minimax version of CER and it has multiple rounds of optimizations to find the best counterfactual perturbations and keeps improving the ranking decision boundaries constantly. As a direct comparison, CAR also generates perturbations and retrains the ranking model on top. Nevertheless, the objective of CAR is to directly optimize the ranking performance. With more near-boundary perturbation examples generated, the embedded feature space will be further covered for the scenarios where the data set is extremely sparse. Similar phenomena have been observed in multiple recent works that use adversarial augmentation~\cite{APR_SIGIR_2018,PURE_KDD_2021,CFGAN_CIKM_2018}. For the attention-based model NAR, due to its high model complexity, it does not perform well for small-scaled data sets because of model overfitting. But its performance gets significantly improved on large-scale data sets and becomes very competitive on them. The VBPR and NCF models in general are quite stable in terms of their performance, but due to the lack of a module for explanation, their prediction results are hardly interpretable.

\begin{figure*}[!t]
\begin{tabular}{ccccc}
\begin{subfigure}{0.21\textwidth} \hspace{-3mm} \includegraphics[width=1\columnwidth]{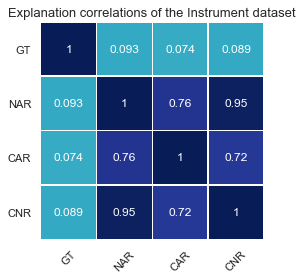}\end{subfigure}&
\begin{subfigure}{0.2\textwidth} \hspace{-6mm} \includegraphics[width=1\columnwidth]{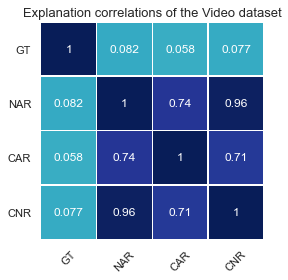}\end{subfigure}&
\begin{subfigure}{0.2\textwidth} \hspace{-9mm} \includegraphics[width=1\columnwidth]{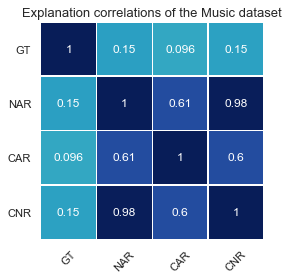}\end{subfigure}&
\begin{subfigure}{0.2\textwidth} \hspace{-12mm} \includegraphics[width=1\columnwidth]{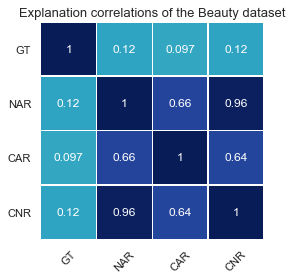}\end{subfigure}& 
\begin{subfigure}{0.2\textwidth} \hspace{-15mm} \includegraphics[width=1\columnwidth]{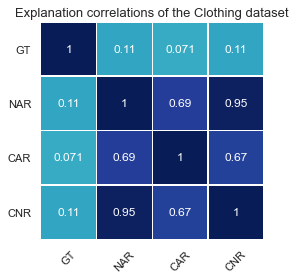}\end{subfigure}\\ 
\end{tabular}
\vspace{-3mm}
\caption{Correlation comparison of different explainable models.}
\label{vis_plot:correlation}
\vspace{-3mm}
\end{figure*}

\begin{figure*}[!t]
\begin{tabular}{ccccc}
\begin{subfigure}{0.21\textwidth} \hspace{-4mm} \includegraphics[width=1\columnwidth]{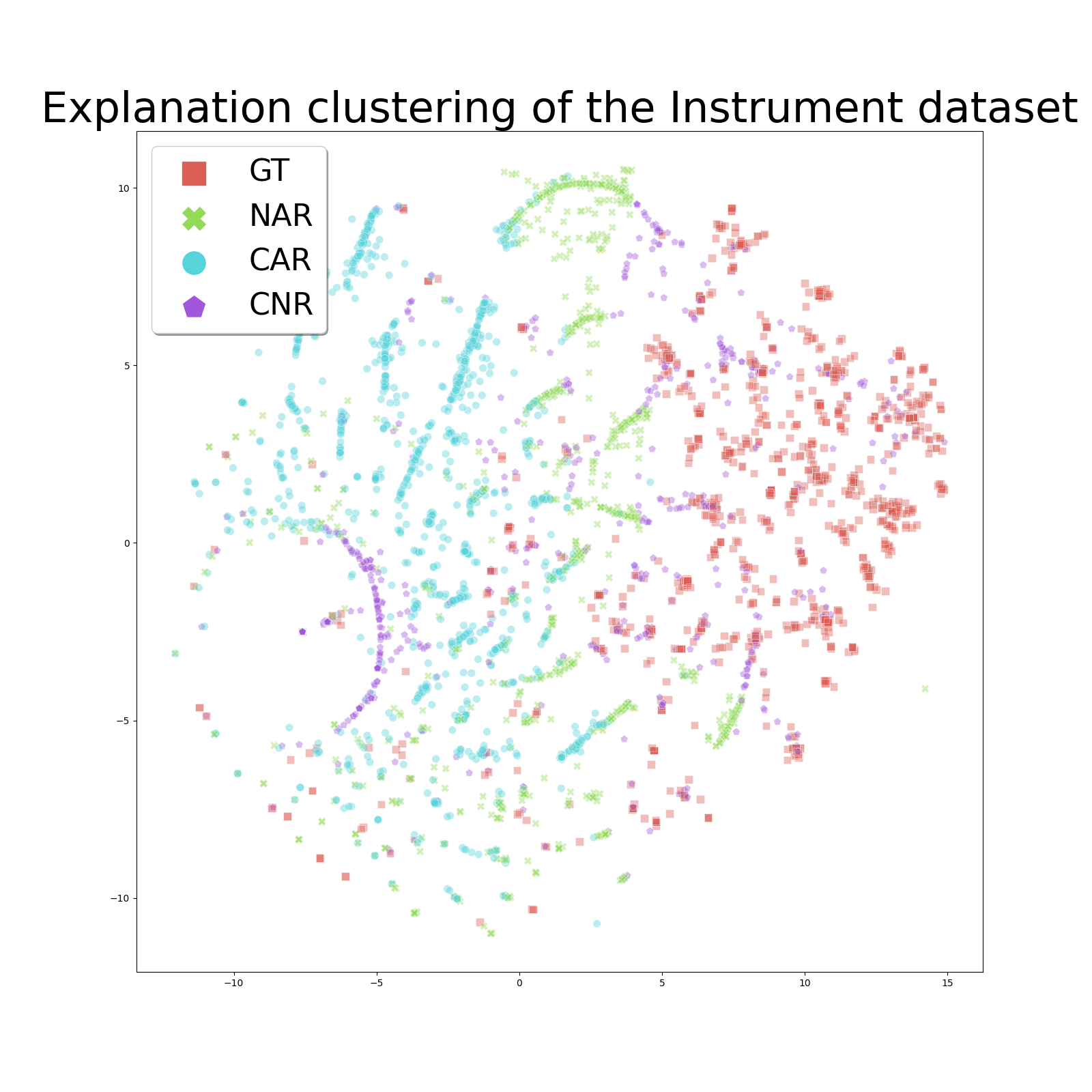}\end{subfigure}&
\begin{subfigure}{0.21\textwidth} \hspace{-8mm} \includegraphics[width=1\columnwidth]{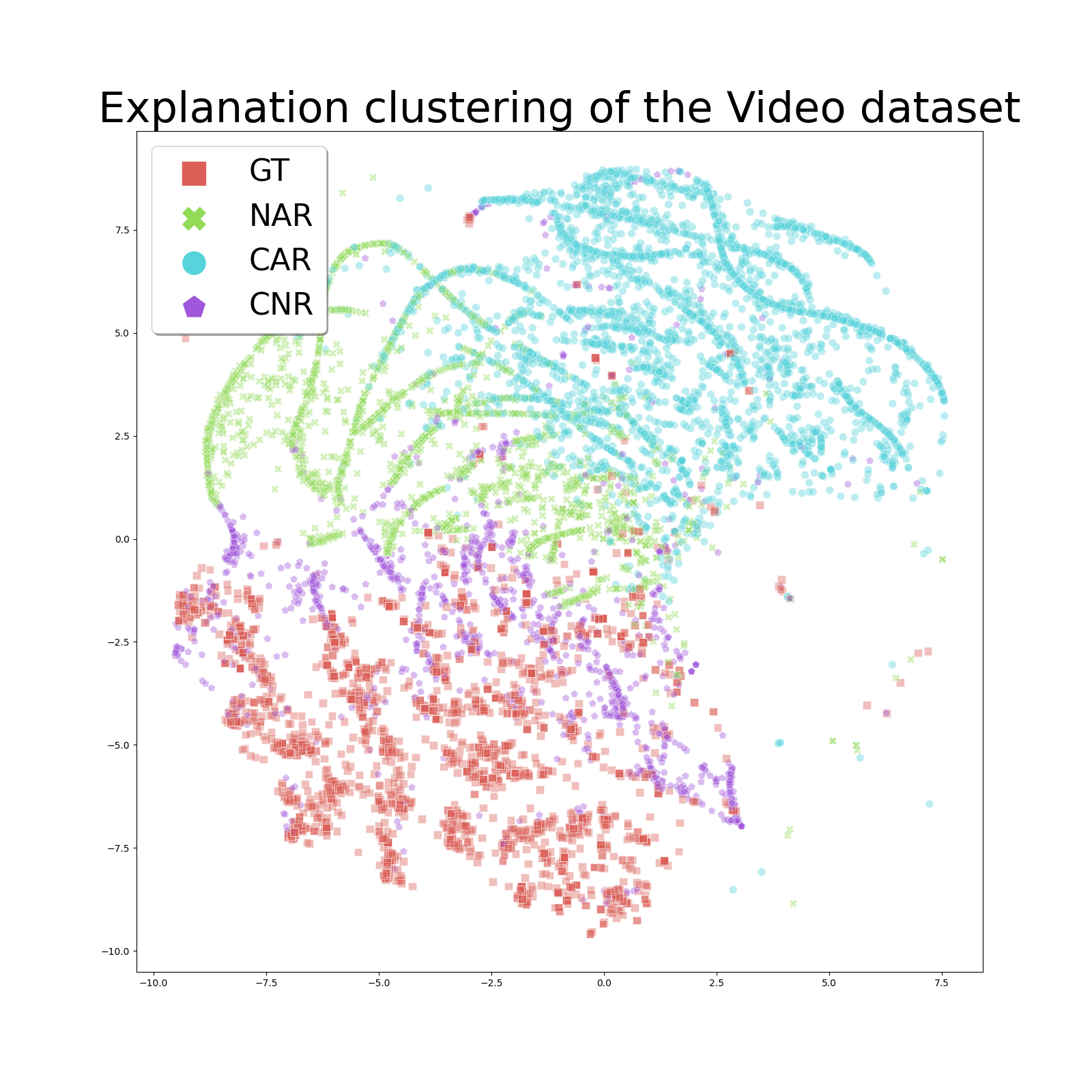}\end{subfigure}&
\begin{subfigure}{0.21\textwidth} \hspace{-12mm} \includegraphics[width=1\columnwidth]{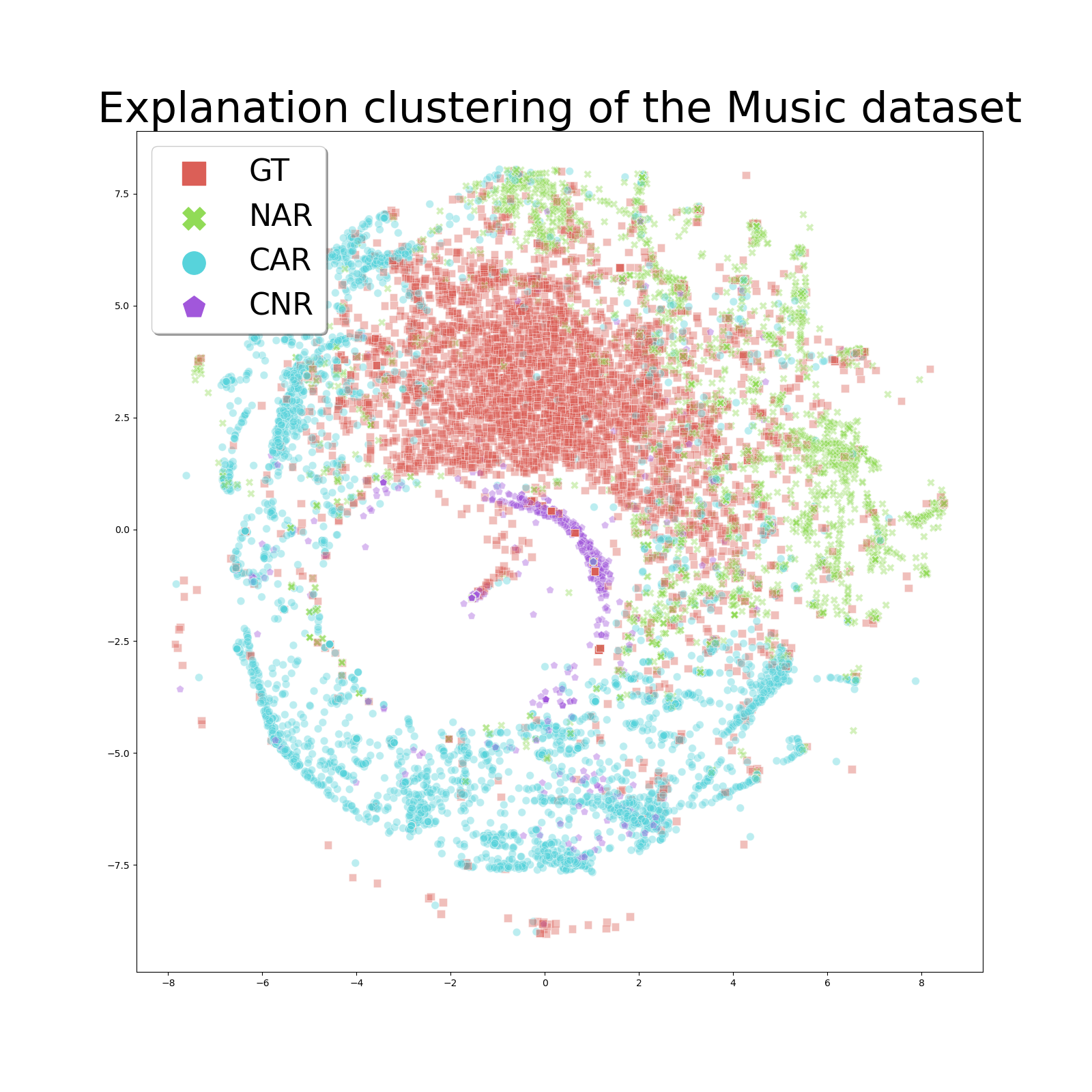}\end{subfigure}&
\begin{subfigure}{0.21\textwidth} \hspace{-16mm} \includegraphics[width=1\columnwidth]{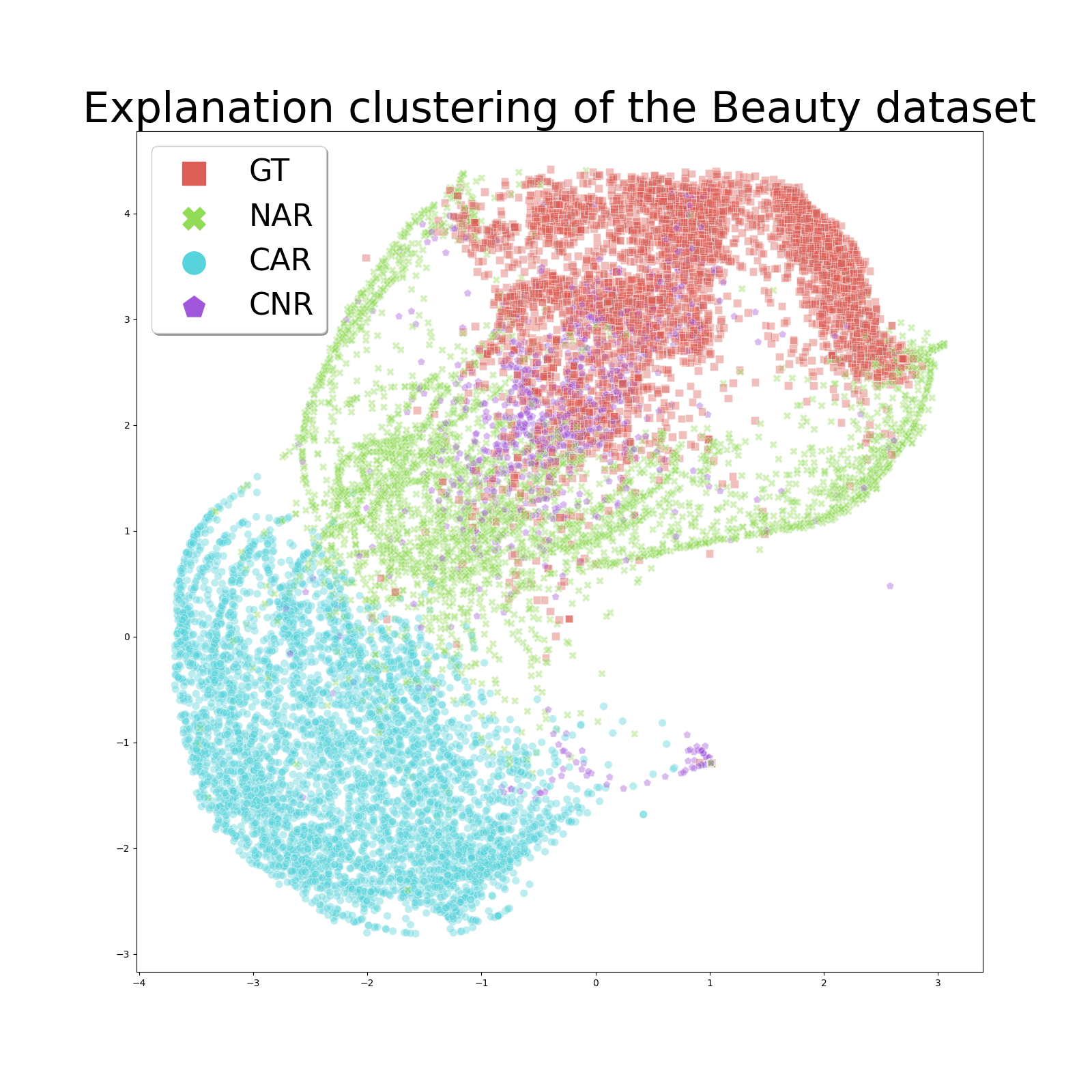}\end{subfigure}& 
\begin{subfigure}{0.21\textwidth} \hspace{-20mm} \includegraphics[width=1\columnwidth]{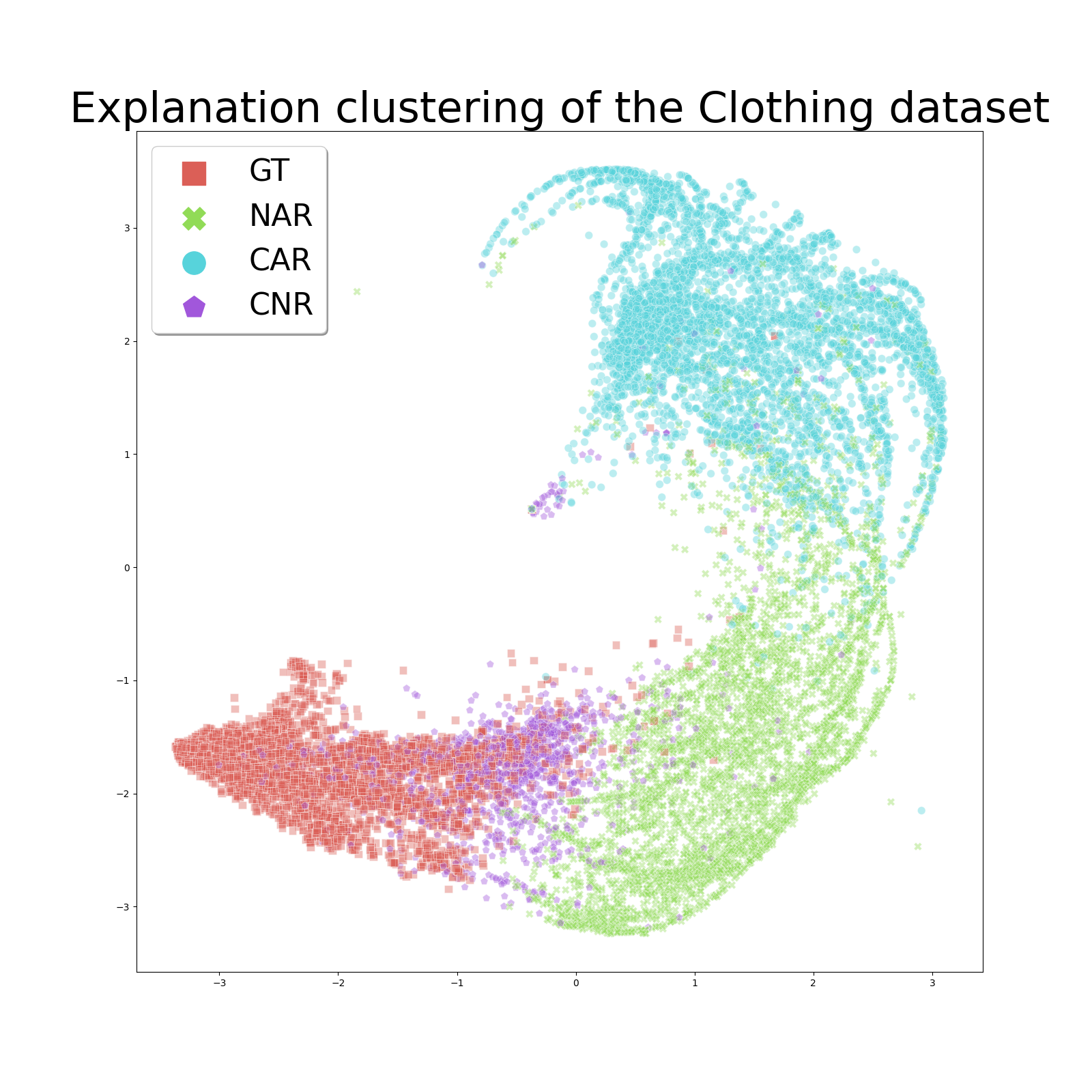}\end{subfigure}\\ 
\end{tabular}
\vspace{-3mm}
\caption{Clustering visualizations of different explainable models.}
\label{vis_plot:cluster}
% \vspace{-3mm}
\end{figure*}

\begin{table*}[!t]
\centering
\setlength\tabcolsep{3pt} % default value: 6pt
\begin{tabular}{l|p{31mm}|p{31mm}|p{31mm}|p{31mm}|p{31mm}}
\toprule
& \textbf{Instrument} & \textbf{Video} & \textbf{Music} & \textbf{Beauty} & \textbf{Clothing} \\ \hline
\textbf{GT}  &  \{vocal, tone, display, pedal, sound\}                               & \{character, drama, season, performances, adventures\}                           & \{hits, artist, performance, playing, charm\}                           & \{hair, pumps, frizzy, scents, salon\} & \{size, material, arch support, strap, velcro\}         \\ \hline
\textbf{NAR} & \{tune, hum, tab, tube, styles\}  & \{effing, murder mystery, outdoors, storytelling, satire\} & \{lyricist, outtakes, comeback, hook, studio\}                      &  \{curler, tingle, cover, itch, pore\}  & \{dressy, cushion, knee, fit, calf\}                \\ \hline
\textbf{CAR} & \{processor, bucks, neck, overdrive, plugs\}                                & \{performance, fans, cabin, twists, performer\}        & \{york, grooves, riffs, intro, sounds\}                    & \{smell, iron, scars, chip, skin tone\}  &  \{fits, boxers, gloves, lining, sleeve\}                \\ \hline
\textbf{CNR} &  \{noise, epiphone, cord, holder, pickup\}                               & \{adventures, seasons, artistry, dramas, series\}         &  \{album, smash, guitar playing, guitar work, songwriters\}                  &  \{hair, drugstore, pencil, perfume smell, customer\}  &   \{bikini, thickness, heels, footbed, padding\}              \\ 
\bottomrule
\end{tabular}
\caption{Top-$k$ words of explanations}
\label{top-k-words-visualization}
\vspace{-4mm}
\end{table*}

\vspace{-2mm}
\subsection{Explanation Evaluation}
In terms of the user-orientated evaluation w.r.t. the explanations, we mainly focus on horizontally comparing all three proposed models. As shown in Figure~\ref{results:visual-evaluation}, we observe that neural attention model NAR and counterfactual model CNR perform relatively well, but the adversarial perturbation model CAR has low scores across all five data sets. The reason is that NAR and CNR are more suitable for model explanations: NAR has an explicitly designed architecture for learning the attention weights of the input features; CNR uses the counterfactual perturbation to alter the ranking decision, therefore, these perturbations are also explainable in terms of making a ranking decision. As a sharp contrast, CAR learns its adversarial perturbations to maximize the ranking loss and they have weak correlations with explanation because these perturbations do not guarantee to change the model ranking. 

\vspace{-2mm}
\subsection{Visualizations Results}
In this section, we compare and connect all three explanation models, NAR, CAR, and CNR, by conducting various visualizations. These visualization show that quantitative evaluations and qualitative visualization are well aligned. Since the users are inclined to give reviews to items with positive or negative sentiment, then, the word-sentiment pairs of each user are one reliable proxy of ground truth (GT) for explanation evaluations.

\subsubsection{Explanation correlation analysis}
In Figure~\ref{vis_plot:correlation}, we compute the average Pearson correlations of GT, NAR, CAR, and CNR in a pairwise manner and demonstrate them as the heatmaps. We observe that comparing GT, NAR and CNR have the most correlated explanations. Since the feature dimension is relatively high, and GT features are extremely sparse (only a few explanation words per user), therefore, the correlation computed w.r.t. GT are in general low as expected. However, the correlations between the proposed explanation models are relatively high. NAR and CNR are also strongly correlated with each other.

\subsubsection{Explanation clustering analysis}
In Figure~\ref{vis_plot:cluster}, we transform all the explanation vectors of users into two-dimensional representations using t-SNE~\cite{van2008visualizing_tsne}. We observe that the users group together within the same model. CNR usually has a cluster of users that are close to the GT user clusters. The NAR clusters are slightly further distanced compared with CNR. CAR always has the most distanced group of users.

\subsubsection{Top explainable words}
As we can see in Table~\ref{top-k-words-visualization}, we randomly pick the users for visualization and observe that the top-5 words of various explanation models are quite different. Nevertheless, they belong to the specific application domains as expected. In general, the retrieved explainable words for all three models also have large variations compared with GT. However, since we are only focusing on the top words, with large $k$ values being selected, we begin to see more overlaps of explainable words.

\vspace{-3mm}
\section{Conclusion}
In this paper, we propose three explainable recommendation models with various transparencies and perform a systematic and extensive comparison between them in terms of both quantitative evaluations and qualitative visualizations. The overall takeaway message for all these proposed models is: CAR has outstanding recommendation performance but does not have sufficient ability to explain; NAR explains well but could easily overfit in small-scale data sets and has reasonably recommendation results on large-scale data sets; CNR is the winner model which balances well in terms of both recommendation performance and explanation. Besides those, CNR is also a blackbox explainable ranking model and it can be easily extended to any existing ranking architecture.


% \begin{acks}
% \end{acks}

\bibliographystyle{ACM-Reference-Format}
\bibliography{acmart.bib}

% \appendix









\end{document}